\documentclass[11pt,a4paper]{article}
\usepackage{arxiv}
\usepackage[utf8]{inputenc} 
\usepackage[T1]{fontenc}    
\usepackage{url}            
\usepackage{booktabs}       
\usepackage{amsfonts}       
\usepackage{nicefrac}       
\usepackage{microtype}      
\usepackage{lipsum}		
\usepackage{graphicx}
\usepackage{doi}
\usepackage{chngcntr}
\usepackage{multirow}
\usepackage{color,soul}
\usepackage{setspace}
\title{Photothermal properties of stable aggregates of gold nanorods}

\date{} 					

\author{Dheeraj Pratap$^{a, }$\thanks{Corresponding author's email: dheeraj.pratap@csio.res.in}, Vikas$^{a, b}$, Rizul Gautam$^{a, b}$, Amit Kumar Shaw$^{a, b}$, Sanjeev Soni$^{a, b}$ \\ \\
	$^{a}$Biomedical Applications Group, CSIR-Central Scientific Instruments Organization, Sector-30C, Chandigarh-160030, India\\
	$^{b}$Academy of Scientific and Innovative Research (AcSIR), Ghaziabad-201002, India\\
}

\renewcommand{\headeright}{}
\renewcommand{\undertitle}{}

\begin{document}
\maketitle

\begin{abstract}
Aggregation of the nanoparticles is a natural phenomena due to various biological and physical parameters but these aggregates are highly unstable. If aggregates would have been stable then it might be useful for some biomedical applications. To study the optical properties and photothermal heat generation of stable aggregates of the metal nanopartilces, we present a forced synthesis of aggregates of small gold nanorods in a mixture of Dulbecco's Modified Eagle's medium (DMEM) and Bovine Serum Albumin (BSA). We synthesised hexadecyltrimethylammonium bromide (CTAB) coated gold nanorods and then prepared the stable aggregates to study the optical characteristics of these aggregates. Absorption spectra of aggregates show the redshift compared to the monodispersive gold nanorods and confirm that the shape and size of aggregate depend on the amount of BSA in DMEM as well as on the concentration of CTAB in the stock solution of the gold nanorods suspension. The higher concentration of the BSA in DMEM and lower concentration of CTAB of monodispersive gold nanorod suspension causes greater redshift. We tuned the well defined plasmonic absorption resonance peaks of the aggregates of gold nanorods up to the second biological therapeutic window. We studied the stability of the synthesised aggregates of gold nanorods for up to one week and found that the aggregates were stable for atleast one week. A photothermal study of these aggregates was also carried out using high power broadband near-infrared light source. Aggregates were also photothermally stable which warrants their repeated use for such applications. The photothermal conversion efficiency of the these stable gold nanorod aggregates were higher than its monodispersive form of the nanorods. The present study  on the optical and photothermal properties of stable aggregates could be useful for biomedical therapeutic applications, sensing, deep tissue light penetration for imaging and other scientific applications to redshift the plasmonic absorption resonance peak and enhance the photothermal effect.
\end{abstract}


\section{Introduction}
Over past decade, nanoparticles have been studied for the applications in hyperthermia~\cite{chatterjee2011nanoparticle,moy2017combinatorial}, plasmonic photothermal therapy (PPTT)~\cite{huang2008plasmonic}, dermatology~\cite{abramovits2010applications}, imaging~\cite{abramovits2010applications,song2019gold}, drug delivery~\cite{douglas1987nanoparticles} etc. Light interaction with nanoparticles inside biological medium produces multi-modal phenomena~\cite{huang2011multimodality}. Usually, most of the techniques which utilise the photothermal phenomena are minimally non-invasive~\cite{jaque2014nanoparticles}. For the biological applications, the nanoparticles have been used in the optical therapeutic windows which range from 700 nm - 980 nm (first window) and 1000 nm - 1400 nm (second window)~\cite{jaque2014nanoparticles}. The photothermal applications of nanoparticles in PPTT and imaging have been extensively explored. In most cases, the gold nanoparticle has been of much interest to the research community because it possesses strong optical absorption resonance at near-infrared (NIR) light, chemical or biological inertness, low cytotoxicity, easy conjugation to a specific antibody or targeting ligand~\cite{baffou2017thermoplasmonics}. In general, the photothermal applications of the nanoparticles are only limited to the first therapeutic window and recently being explored a lot for second therapeutic window. This might be either due to difficulty to get smaller size nanoparticles that have absorption resonance in the second window. For the \textit{in vivo} use, the size of nanoparticles is a crucial factor and it should be small about 50 nm or preferably even less~\cite{soni2015experimental}. There are only a very few types of nanostructures as gold nanoring~\cite{larsson2007sensing}, double gold nanopillars with silica spacer~\cite{kubo2011double}, gold nanocrescent antenna~\cite{bukasov2010probing}, gold nanoshell with silica core~\cite{oldenburg1998nanoengineering}, gold nanodumble~\cite{grzelczak2008influence}, and gold nanorod-in-shell with silver spacer~\cite{tsai2013nanorod} whose localised surface plasmon resonance (LSPR) lies in the second therapeutic window preventing practical feasibility for \textit{in vivo} use because either they are made on the substrate or through complex fabrication processes. Most of the second therapeutic window responsive nanoparticles are made of bi-materials means that they constitute not only gold but also silver, copper or silica which might induce toxicity during the \textit{in vivo} uses. In a simple case of the gold nanorod, by varying the aspect ratio, one can easily get LSPR resonance in the second therapeutic window, however, the longitudinal size would be too large but it should be limited. Jain et al. showed that the calculated LSPR absorption efficiency of the gold nanorod is much higher than the nanosphere or nanoshell~\cite{jain2006plasmon}. Even the synthesis of the gold nanorod is easier as compared to the complicated structure of the nanoshell or any bi-material nanoparticle~\cite{cai2008applications,tsai2013nanorod}. Numerically it had been reported that head-to-head coupling of the spheroids and nanorods might cause a significant redshift of the LSPR resonance wavelength~\cite{jain2008surface}. Practically, clustering or aggregating the small nanorods might be a good choice to get the desired resonance in the second therapeutic window. The tuning of the localised surface plasmon resonance (LSPR) at a higher wavelength is one quest for the biomedical application especially for the plasmonic photothermal therapy (PPTT)~\cite{dickerson2008gold,wang2014synergistic}. Gu et al. reported that if we increase the concentration of gold nanorods in its monodispersive form then there was increased plasmonic coupling and very small redshift in the plasmonic resonance~\cite{gu2018effect}. Chen et al. showed the PPTT of a single cell where gold nanorods get naturally accumulated within the cell~\cite{chen2017photothermal}. Surprisingly, for the PPTT they used the laser whose wavelength was near to the resonance of monodispersive nanorod but for accumulated nanorods, the resonance would be quite different than the monodispersive form. Zook et al. reported that using the Dulbecco's Modified Eagle's medium (DMEM) and Bovine Serum Albumin (BSA) spherical gold, silver, cerium oxide and polystyrene nanoparticles could be converted into stable aggregates or agglomerate form for a minimum of two days but they did not provide any structural characterization with the scanning electron microscope (SEM) or tunnelling electron microscope (TEM) and not even the photothermal study of the aggregates~\cite{zook2011stable}. In our report we present the details of the optical and photothermal study of aggregates of gold nanorods which are stable for atleast one week. 

In this article, we report the synthesis of stable aggregates of gold nanorods, their structural and optical properties, and photothermal effects. The organization of the paper is in the following manner. Section-2 describes the details of materials and methods of gold nanorods and their aggregates synthesis and characterizations. In section-3, we analyse and discuss the optical properties and photothermal effects. The last section-4 concludes the work and gives an insight into possible applications. 
\section{Materials and methods}
\begin{figure}[h]
\centering
\includegraphics[width=1\textwidth]{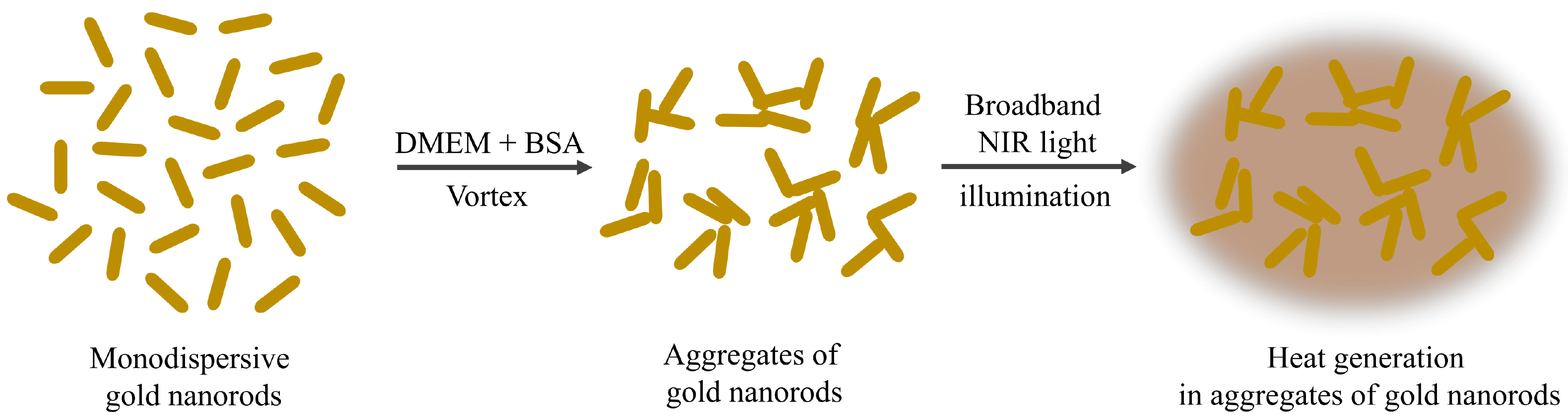}
\caption{Schematic of the aggregation of gold nanorods in a mixture of Dulbecco's Modified Eagle's medium (DMEM) and Bovine Serum Albumin (BSA), and the photothermal effect in the gold nanorods aggregates after illuminating with broadband near infra-red (NIR) light.}
\label{fig:schematic}
\end{figure}
A brief theme of the experimental work is delineated in Fig.~\ref{fig:schematic}. In the experiment, it should be noted that the total number of gold nanorods were the same (approximately) in monodispersive form as well as in their corresponding aggregates form by keeping the same concentration of the gold nanorods in each set. The desired gold nanorods were successfully synthesized using a seed-mediated method with slight variation as protocol mentioned in Ref.~\cite{jia2015synthesis,chang2018mini}. This protocol also shows the repeatability of synthesized gold nanorods. For the gold nanorod synthesis, we obtained chloroauric acid (HAuCl$_4$), silver nitrate (AgNO$_3$), cetyltrimethylammonium bromide (CTAB) and sodium borohydrate (NaBH$_4$) from Sigma-Aldrich, and ascorbic acid from Sisco Research Laboratories and hydrochloric acid (HCl) from Finar. At the first step, a 2 ml (0.01 M) solution of NaBH$_4$ was prepared and kept in the freezer to be used as ice-cold. An amount of 0.25 ml (0.01 M) of HAuCl$_4$ solution was added to 9.75 ml (0.1 M) CTAB solution under vigorous stirring. Followed by 0.6 ml (0.01M) freshly prepared ice-cold NaBH$_4$ was added to the mixture of HAuCl$_4$ and CTAB mixture under vigorous stirring for two minutes. The resultant solution became yellowish-brown and was kept two hours at ambient conditions before use. This was the seed solution for the nanorod synthesis. The growth solution was obtained by sequentially adding 5 ml (0.01 M) HAuCl$_4$, 1 ml (0.01 M) AgNO$_3$, and 2 ml (1.0 M) HCl  to 90 ml (0.1 M) CTAB  solution under gentle stirring and followed by adding a 0.8 ml (0.1 M) fresh ascorbic acid solution under rapid stirring up to solution become transparent. The whole amount of seed solution was added to the growth solution under vigorous stirring for two minutes and then the final mixture was left overnight at 35~$^{\circ}$C for the growth of the nanorods. On the next day, the whole solution containing the nanorods were centrifuged by dividing it into an equal amount in four round-bottom 50 ml centrifuge tubes at 15000 rpm and 30 $^\circ$C for 20 minutes using a Remi-C24 plus centrifuge. The excess amount of the CTAB was collected as a supernatant and then discarded. The synthesized gold nanorods obtained in the pellet were further washed once in DI water at the same conditions and then dispersed in water. Since the amount of CTAB was extremely large compared to other components used for gold nanorod synthesis, therefore, the amount of pellet and added DI water were noted carefully to calculate the  concentration of CTAB in the final nanorod suspension. The total volume of the final solution was 24~ml and kept as stock suspension of nanorods for further uses which had a measured concentration of 15.9 mg/l using the inductively coupled plasma mass spectrometry (ICP-MS). The structural characterization was carried out with a transmission electron microscope (TEM) of JEOL JEM-2100CR. 

To transform gold nanorods into aggregates we used phenol red free Dulbecco’s Modified Eagle Medium (DMEM) from Sigma-Aldrich and Bovine Serum Albumin (BSA) from Sisco Research Laboratories. We prepared two solutions 2\% (w/v) and 4\% (w/v) by dissolving BSA in DMEM. Further 1:9 diluted sample was prepared (with 100~$\mu$l stock suspension and 900~$\mu$l DI water) so that it becomes completely monodispersive form and there should no plasmonic coupling exist. Visual appearance of the diluted nanorods suspension was a very faint pink color. The diluted gold nanorods suspension was used as a reference concerning aggregates. The calculated CTAB concentration in diluted gold nanorod suspension was 0.42~mM and we note this concentration as C1 (see the supporting information). A similar method to calculate the concentration of CTAB has been used by Wu et al.~\cite{wu2015large}. We added 100 $\mu$l stock solution into 100 $\mu$l DI water and this diluted solution was added to 800 $\mu$l each 2\% and 4\% BSA-DMEM solution and vortexed at 1900 rpm for 5 minutes. The aggregate was visually appeared like faint brick color. After vortexing, the absorption spectra of sample was recorded. In the first set, we had three samples with C1 concentration of CTAB, first diluted gold nanorod (C1-NR), second and third diluted gold nanorod mixed with 2\% and 4\% BSA (C1-NR+2\%BSA and C1-NR+4\%BSA). We used a UV-visible spectrometer from LABINDIA UV 3200 to record the absorption spectra of the monodispersive gold nanorods and their aggregates. Absorption spectra were recorded just after vortexing the gold nanorods in BSA-DMEM solutions. To check the stability of the aggregates we took its absorption spectra after 3 hours, 72 hours and 144 hours after vortexing. The Zeta-potential measurement was carried out by Zetasizer Nano series from Malvern. 

Another set of three samples of gold nanorods and aggregates at a lesser concentration ($<$ C1) of CTAB  were also prepared to check the effect of CTAB on aggregation. To prepare the second set of samples, we centrifuged 2 ml of stock nanorod suspension in eppendorf tube at 17000 rpm and 25~$^{\circ}$C for 20 minutes. The obtained visually colorless and transparent supernatant was then discarded and then the gold nanorods in the pellet were again redispersed in DI water to make the final volume of 2~ml. The amount of supernatant removed and the amount of water added was carefully noted to calculate the approximate concentration of CTAB for the second set of samples. This gold nanorod suspension again diluted for the first set of samples to transform into aggregates and taking absorption spectra. The calculated concentration of CTAB was 0.08~mM in the second set of diluted gold nanorod suspension and we noted  that concentration of CTAB as C2 (see the supporting information). The same procedure as earlier was followed to get nanorods aggregates for 2\% and 4\% BSA in DMEM mixture. Now we have another set of three samples, first monodispersive gold nanorods (C2-NR), second and third aggregates C2-NR+2\%BSA and C2-NR+4\%BSA respectively. 

The photothermal response was measured using a homemade photothermal device. That device having a halogen lamp as a vis-NIR broadband optical source. An optical filter was used to cut off the undesirable spectral range. After passing through a filter, the light was collected through a multi-modal optical fiber bundle of diameter 12 mm. The final output of the light through the fiber bundle has a top-hat shape profile from 720 nm to 870 nm with an irradiance 1.3 W/cm$^2$ which have a maximum value at 795 nm. One such configuration of the photothermal device is reported for visible light source in Ref.~\cite{soni2015experimental}. Total incident power on the sample was 1.8 W. We used 2 ml each of the nanorods and aggregates solutions in a 4 ml quartz cuvette of the square cross-section for photothermal measurement. The temperature of the solution was measured by a ``K" type thermocouple and the obtained output was recorded using the national instruments data acquisition device NI-myDAQ. During the photothermal analysis, the rise in temperature of the solution was recorded for 15 minutes after irradiating the solution with the light source through fiber bundle. The temperature change was estimated in reference to the ambient temperature. For reference, the rise in temperature of deionized water of 18.1~M$\Omega$-cm resistivity was also recorded. The photothermal conversion efficiency (PCE) was calculated at the ambient conditions using the method reported in Ref.~\cite{jiang2013size,alrahili2020morphology} with the broadband light source of the above photothermal device and listed in the Table~\ref{tab:table}.

\section{Results and discussion}
\begin{figure}[h]
\centering
\includegraphics[width=1\textwidth]{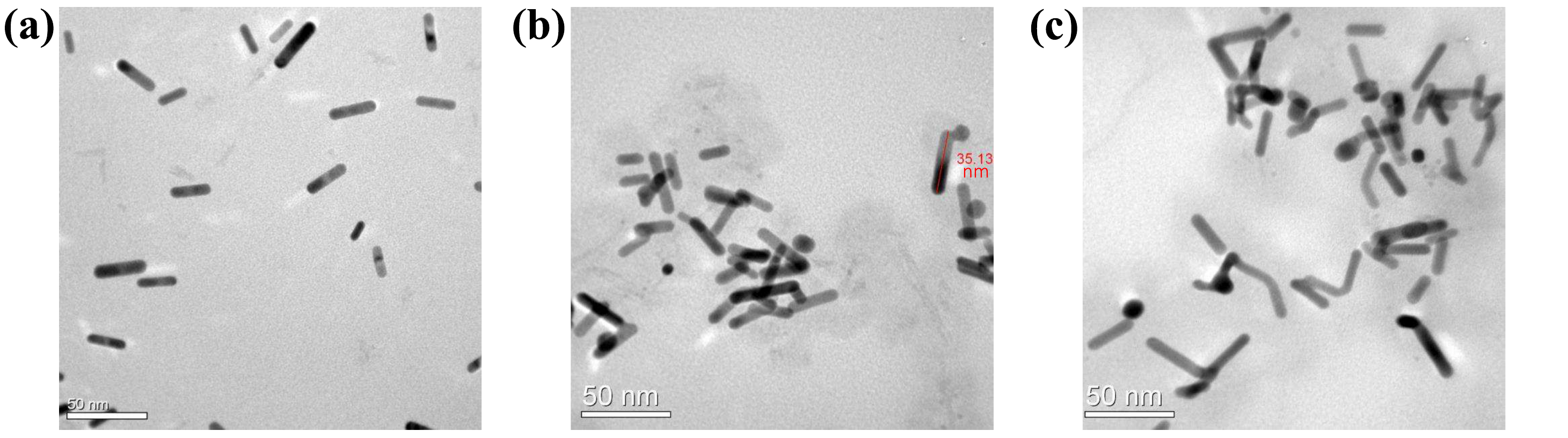}
\caption{Transmission electron microscope (TEM) images of (a) the monodispersive gold nanorods in DI water; and their aggregates in (b) 2\% BSA-DMEM  and (c) in 4\% BSA-DMEM solution at the C1 (3.04 mM) concentration of the hexadecyltrimethylammonium bromide (CTAB). BSA stands for Bovine Serum Albumin and DMEM stands for Dulbecco's Modified Eagle's medium. In each sub-figure the scale bar is 50 nm.}
\label{fig:TEM_image}
\end{figure}
The TEM images of the gold nanorods and their aggregates are shown in Fig.~\ref{fig:TEM_image}. For the TEM imaging, there the same dilution sample was used as for the absorption spectra when the concentration of the CTAB was C1. We can see in Fig.~\ref{fig:TEM_image}(a) that the nanorods appear separate from each other and these are not in contact with each other. This happened due to the very dilute concentration of the monodispersive gold nanorod suspension. Since the nanorods concentration was kept very dilute to minimise the internal interaction with each other, therefore, it was hard to obtain the locations of nanorods on the TEM grid. The average size of the gold nanorod was 24.6 $\pm$ 3.3 nm $\times$ 5.9 $\pm$ 0.1 nm. Figure~\ref{fig:TEM_image}(b) and  Fig~\ref{fig:TEM_image}(c) show aggregates of the gold nanorods in 2\% BSA-DMEM mixture and 4\% BSA-DMEM mixture respectively. We can see that there are no monodispersive forms of the gold nanorods and they are in contact with each other in proximity and making random chain structures. We should note that this is a 2D image of the aggregates which is taken on a copper grid. This image does not give the true information of the orientation of the nanorods in the aggregate because, in the solution of the aggregates, the nanorods should be randomly oriented in 3D. Their orientation and distribution would depend on the surface charges of the aggregates and solution constituents which could be experimentally controlled. When the aggregate solution is cast on a TEM grid and dried for TEM imaging, it becomes an effectively 2D structure. We could not get the true 3D orientation of the nanorods in their aggregates form. The measured Zeta-potentials for the nanorods and aggregates with 2~\%BSA and 4~\%BSA were 24.93~mV, -14.57~mV and -12.08~mV respectively. 

\begin{figure}[h!]
\centering
\includegraphics[width=1\textwidth]{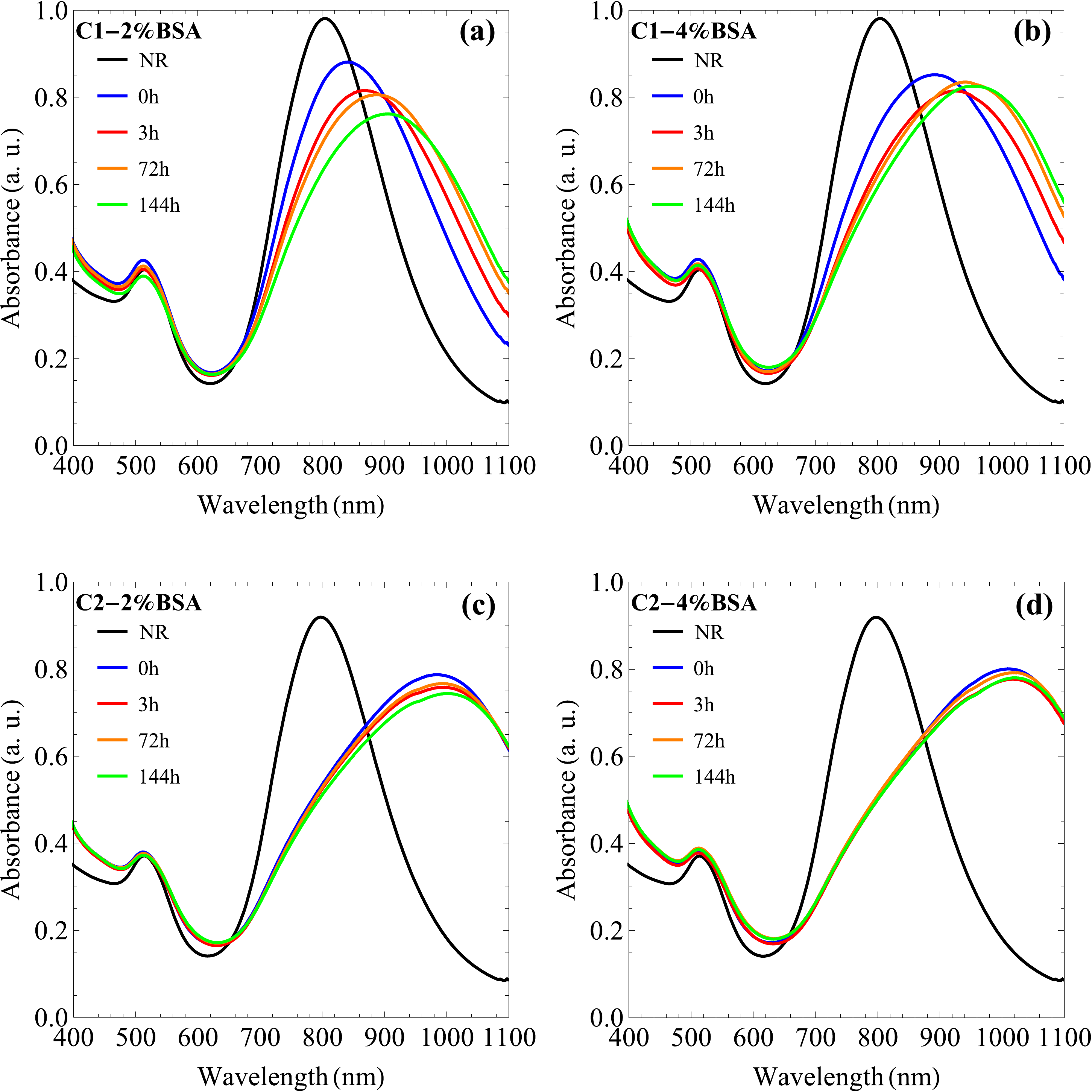}
\caption{Absorption spectra of the nanorods and its aggregates at different concentrations C1 (3.04 mM) and C2 (0.61 mM) of hexadecyltrimethylammonium bromide (CTAB) with time. (a) with 2\% BSA and (b) 4\% BSA for first sample set with CTAB concentration C1. (c) 2\% BSA and (d) 4\% BSA for the second set of samples with CTAB concentration C2. BSA notify Bovine Serum Albumin and DMEM notify the Dulbecco's Modified Eagle's medium.}
\label{fig:C1_C2_2p_4p_BSA}
\end{figure}

The absorption spectra of the gold nanorods and their aggregates are shown in Fig.~\ref{fig:C1_C2_2p_4p_BSA}. For the CTAB concentration C1, the monodispersive gold nanorods have longitudinal and transverse resonance peaks at 803 nm and 514 nm respectively and gold nanorod aggregates have well-defined peaks also but having redshifts concerning the monodispersive gold nanorods. When the concentration of BSA is 2\% then the absorption peaks of the aggregates are around 900 nm and for 4\% BSA peaks are around 950 nm. The strongest absorption LSPR peaks of the gold nanorods and their aggregates are listed in Table~\ref{tab:table}. We can see that increasing the concentration of BSA increases the aggregate size, therefore, there is a higher redshift in the plasmonic absorption resonance~\cite{zook2011stable}. This happens because the BSA neutralizes the surface charge of gold nanorods as well as the extra charge of CTAB in the solution thus binds them together which gives a defined absorption resonance peak. The absorption spectra of aggregate has slight broader peak compared to the monodispersive form of gold nanorods which show there are some variation in the size of the aggregates. As time increases there is a very slight redshift in the resonance peak of the aggregate. If we compare the absorption peaks of aggregates at 0 hour and 3 hours then it could be inferred that to get some proper size of aggregates takes some time.

\begin{figure}[h!]
\centering
\includegraphics[width=0.5\textwidth]{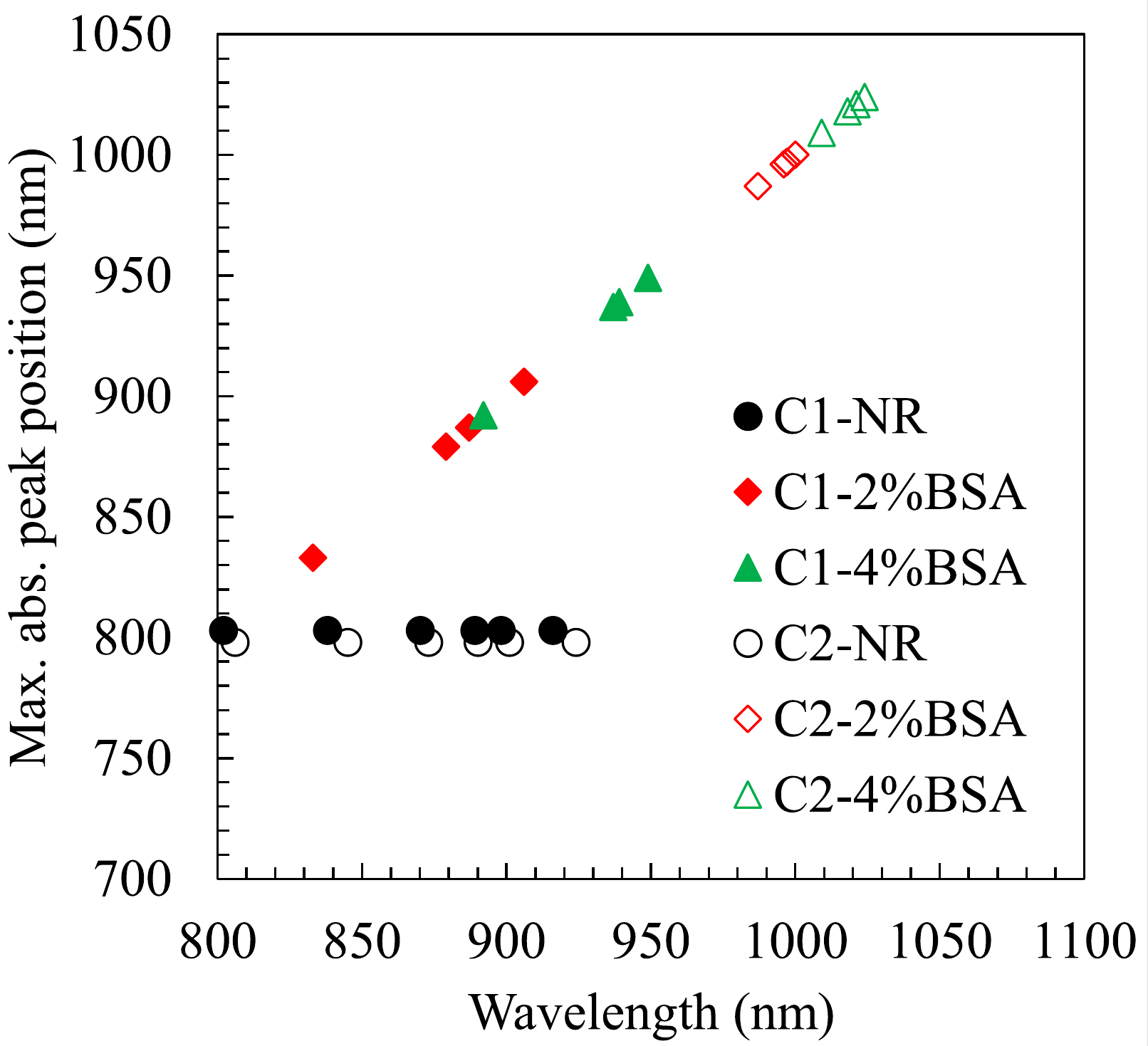}
\caption{The maximum absorption peak position of the gold nanorods and their aggregates with respect to the incident wavelength for time span from 0~h to 144~h. Solid marker: CTAB concentration C1, and hollow marker: CTAB concentration C1.}
\label{fig:lspr_peaks_vs_lda}
\end{figure}

For CTAB concentration C2 ($<$ C1), the gold nanorods have longitudinal and transverse absorption resonance peaks are at 798 nm and 514 nm respectively as shown in Fig.~\ref{fig:C1_C2_2p_4p_BSA}(c) and Fig.~\ref{fig:C1_C2_2p_4p_BSA}(d). The strongest absorption peak wavelengths of the aggregates for the second set of aggregates are listed in Table~\ref{tab:table}. We see that for low concentration C2 of CTAB the transverse peak of the gold nanorod is the same as for concentration C1 but there is a very slight blueshift of 5~nm in the longitudinal resonance peak. For the lower concentration C2 of CTAB, there is a huge redshift in the absorption resonance peaks of the aggregates around 1000 nm and 1020 nm for 2\% and 4\% BSA concentrations respectively. The large redshift in the resonance peak at lower CTAB concentration case is because the BSA neutralises more surface charge of the nanorods and converts into larger aggregates. In earlier case for higher concentration of the CTAB (C1), BSA had to neutralise more CTAB in the solution and fewer surface charges of the gold nanorods and convert into a smaller size of aggregates. In the second set of aggregates, the absorption peaks are more stable than the first set of aggregates. By varying the size of the gold nanorods as well as the the concentration of CTAB and BSA one could control the absorption peak wavelengths of the aggregates. The peak positions for the maximum absorption with respect to the incident wavelength of the gold nanorods and their aggregates synthesized at different concentrations of BSA and CTAB are plotted in Fig.~\ref{fig:lspr_peaks_vs_lda} from the time span 0~h to 144~h. Points with solid markers correspond to the aggregates with a higher concentration of the CTAB (C1) and hollow markers represents the aggregates with a lower concentration of the CTAB (C2). It can be seen that the aggregates were stable. The stability of the aggregates for lower CTAB concentration is better than the stability of the aggregates with higher CTAB concentration.
%
%
\begin{table}
	\caption{Prominent absorption peak wavelengths of the gold nanorod aggregates after 0 h, 3 h, 72 h and 144 h from its synthesis and photothermal conversion efficiency (PCE). Concentration of the hexadecyltrimethylammonium bromide (CTAB), C1 = 3.04 mM and C2 = 0.61 mM. BSA-Bovine Serum Albumin and DMEM-Dulbecco's Modified Eagle's medium.}
	\centering
	\begin{tabular}{lccccc}
		\toprule	 	  
		\multirow{2}{*}{Sample} & \multicolumn{4}{c}{Strongest absorption peak wavelength (nm)} & \multirow{2}{*}{PCE (\%)}\\
		\cmidrule(r){2-5}
		 	   	  & 0 h  & 3 h  & 72 h & 144 h  \\
		\midrule
		C1-NR     & 803  & 803  & 803  & 803  & 65\\ 
		C1-2\%BSA & 833  & 879  & 887  & 906  & 65\\ 
		C1-4\%BSA & 892  & 937  & 939  & 949  & 61\\ 
		C2-NR     & 798  & 798  & 798  & 798  & 60\\ 
		C2-2\%BSA & 987  & 997  & 996  & 1000 & 65\\ 
		C2-4\%BSA & 1009 & 1018 & 1024 & 1021 & 67\\ 
		\bottomrule
	\end{tabular}
	\label{tab:table}
\end{table}

The dipole interaction theory describes the spectral shift in the surface plasmon resonance of the nanoparticles plasmonic coupling~\cite{tsai2012plasmonic,zhang2014plasmon}. In the presence of an electric field, nanoparticle acts as a small dipole. The nanorods have anisotropic plasmonic behaviour. Ideally, when nanorods are aligned side-by-side to each then all the dipoles are oriented in the same direction and feel a repulsive force (increases the energy of the system) and combined resonance would have a blueshift while nanorods are aligned end-to-end then dipoles feel the attractive force (decreased energy of the system) and combined resonance have a redshift. As the number of nanorods increases then there would be a larger spectral shift.  As experimentally observed that aggregates show redshift, therefore, in the suspension the end-to-end nanorods coupling dominates and gives a large redshift. When the concentration of the BSA increases then the number of nanorods in aggregates increases and gives a larger redshift.

\begin{figure}[h!]
\centering
\includegraphics[width=1\textwidth]{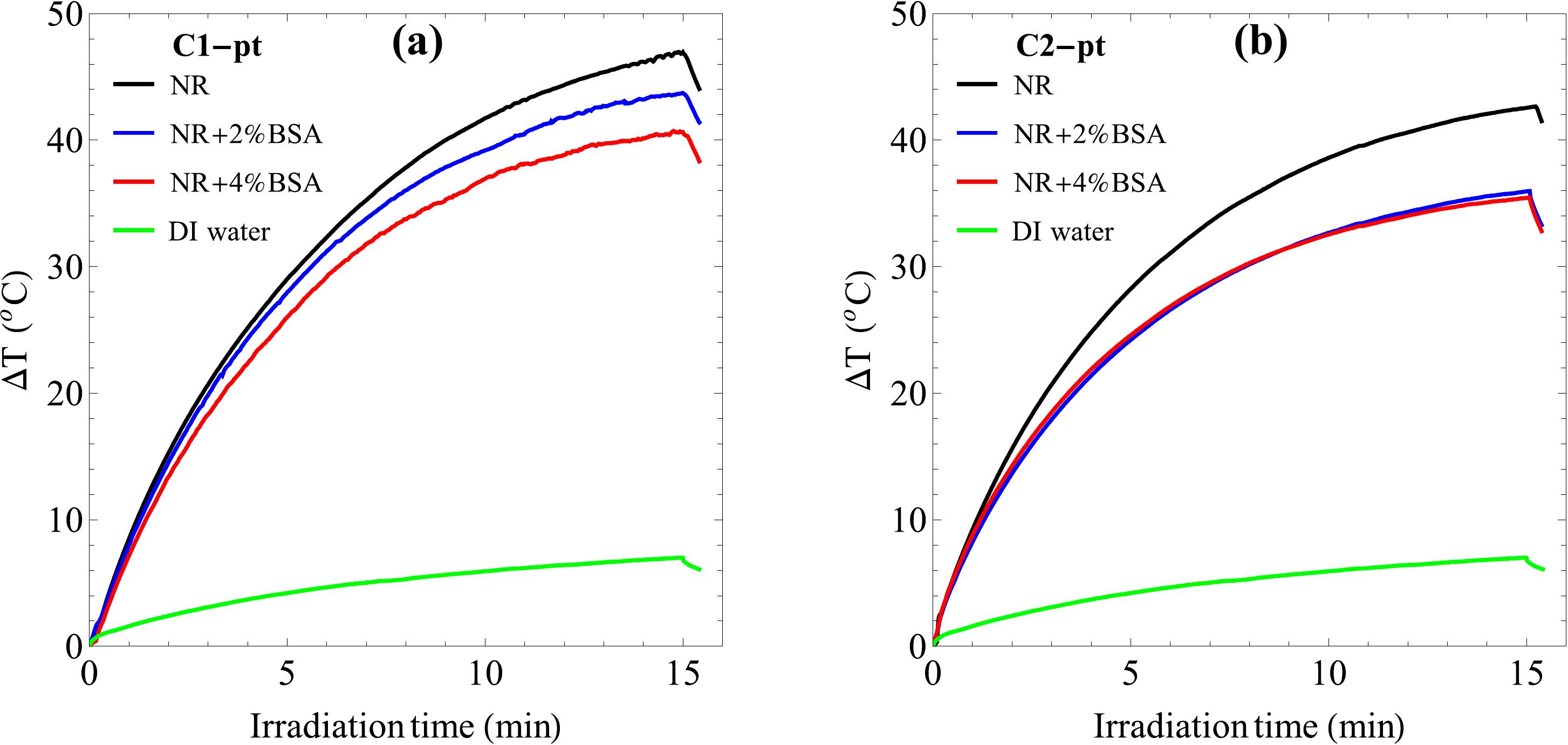}
\caption{Temperature evolution in the gold nanorods and aggregates for hexadecyltrimethylammonium bromide (CTAB) concentration (a) C1 = 3.04 mM and (b) C2 = 0.61 mM at different 2\% and 4\% BSA concentration after the photothermal (pt) process irradiating with NIR light for 15 minutes. Initial temperature, $T_0 = 30^{\circ} \mathrm{C}$. BSA-Bovine Serum Albumin and DMEM-Dulbecco's Modified Eagle's medium.}
\label{fig:C1_C2_ppt_min}
\end{figure}

In the photothermal phenomena, when light is absorbed by the metal nanoparticle then absorbed energy is converted into heat. Irradiating the gold nanorods and their aggregates with the NIR broadband light, it was found that temperature increases over time and approaches towards saturation that is the stationary state condition. Figure~\ref{fig:C1_C2_ppt_min} shows the temperature evolution in the gold nanorods and their aggregates for CTAB concentration C1 and C2. Figure~\ref{fig:C1_C2_ppt_min}(a) shows the increased temperature of the gold nanorod solution and aggregates solution at CTAB concentration C1 for 2\% and 4\% BSA in DMEM. The photothermal efficiency depends on the total amount of absorbed and scattered energies~\cite{baffou2013thermo}. Since the incident light source has maximum power at 795 nm and decreases either side, therefore, the monodispersive nanorods have the highest temperature increase. Temperature rise for the aggregate of 2\% BSA is higher than the aggregate of 4\% BSA because the absorption peak of later is the farthest right side from the maximum peak position of the light source spectrum. We used a broadband light source that's why this temperature increase happens in aggregates otherwise if we would have used a laser source of wavelength 795 nm then there would be not much rise in temperature for the aggregates. For comparison, the temperature rise of DI water also shown in the plot which shows temperature rise only about 7 $^\circ$C. Figure~\ref{fig:C1_C2_ppt_min}(b) shows the photothermal effect of gold nanorods and aggregates for lower concentration C2 of the CTAB. As we have noticed that in the case of the second set of samples, absorption peak wavelength positions for 2\% and 4\% BSA aggregates are very close to each other (Fig.~\ref{fig:C1_C2_2p_4p_BSA}(c-d)) and very far from the incident light spectral range, therefore, in both case the temperature rise is very close to each other. The PCE listed in Table~\ref{tab:table} for the aggregate C1-2\%BSA (65\%) was equivalent and lower for C1-4\%BSA (61\%) in comparison to the monodispersive nanorods C1-NR (65\%) while it was higher for C2-2\%BSA (65\%) and C2-4\%BSA (67\%) in comparison to C2-NR (60\%). The PCE of aggregates should be higher with respect to its monodispersive nanorods as the numerical results showed higher heat generation in aggregates due to enhanced plasmonic coupling within the nanorods~\cite{pratap2021photothermal}. Discrepancies in the PCE for sample C1-4\%BSA might be due to the experimental errors.

\begin{figure}[h!]
\centering
\includegraphics[width=1\textwidth]{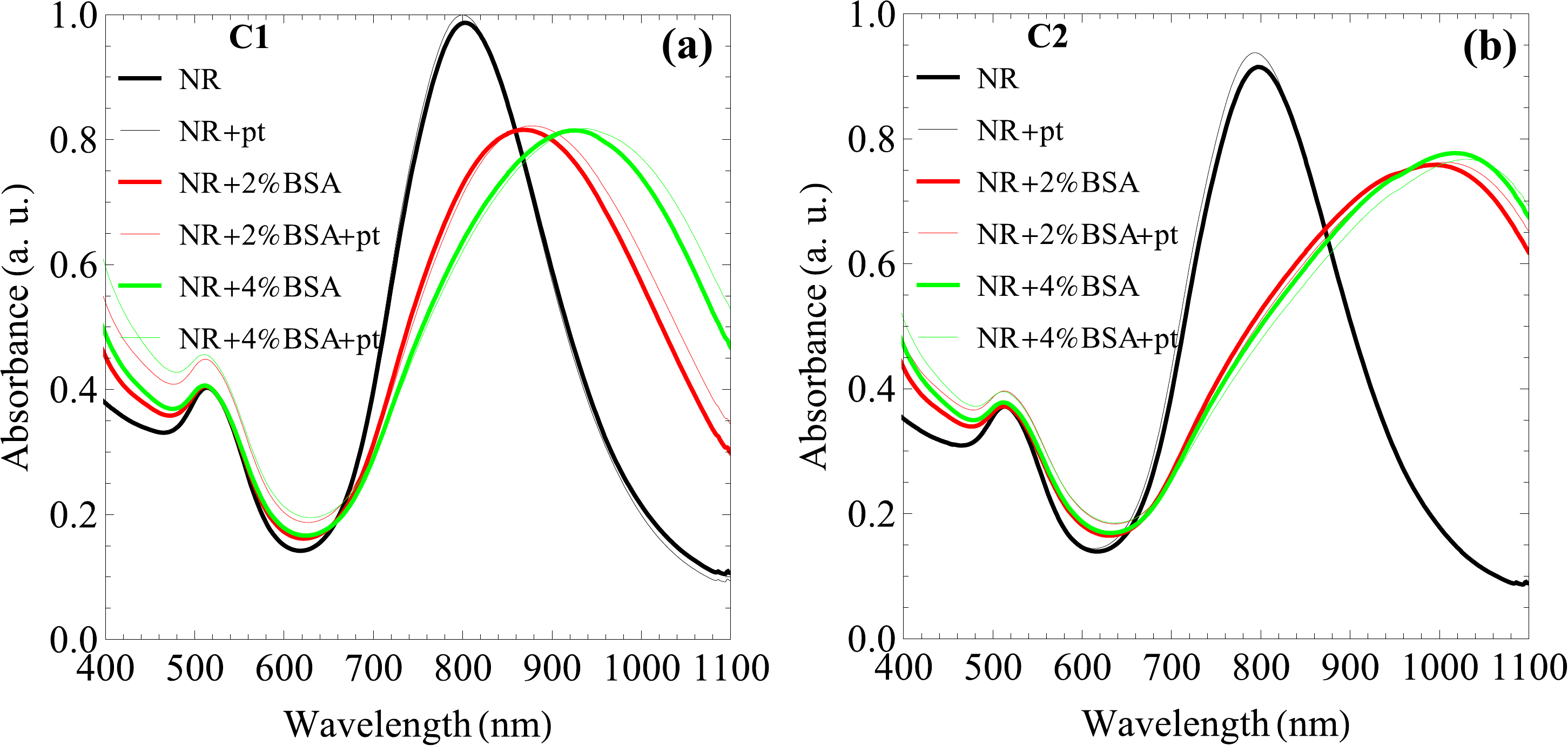}
\caption{Absorption of the gold nanorods and aggregates before (thick lines) and after (thin lines) the photothermal (pt) process irradiating for 15 minutes for the hexadecyltrimethylammonium bromide (CTAB) concentration (a) C1 =3.04 mM and (b) C2 = 0.61 mM. BSA-Bovine Serum Albumin and DMEM-Dulbecco's Modified Eagle's medium.}
\label{fig:C1_C2_ppt_shift}
\end{figure}
We used the broadband light source at a power of 1.8 W to study the photothermal effects in the gold nanorods for its monodispersed and aggregate forms. We had studied the stability of the gold nanorods and aggregates also after the photothermal interaction by measuring  their absorption spectra. Figure~\ref{fig:C1_C2_ppt_shift} shows the absorption spectrum of the gold nanorods and their aggregates before (thick line) and after (thin line) the photothermal process for both sets of samples when the concentration of the CTAB was C1 and C2. We can see that the gold nanorods and aggregates have almost the same absorption peak wavelengths before and after the photothermal measurement. There is a very slight change in the peak positions. Therefore, at a power 1.8 W broadband NIR light source the nanorods and aggregates could be reused for photothermal conversion applications. At a very high power pulsed and continuous laser source it had been shown that the gold nanoparticles might reshape~\cite{fales2017quantitative,harris2017gold}. Sometimes, the reshaping of the non-spherical particles might be a threat for plasmonic photothermal applications. Therefore, it would be better to use a sufficient broadband light source whose spectrum could cover the absorption resonance wavelength of the nanoparticles even if there is slight change in the shape of the nanoparticles. 
\section{Conclusions}
It is concluded that the stable aggregation of the gold nanorods could be obtained by introducing the BSA and DMEM environment to the nanorods suspension. The absorption resonance peak of aggregate did not depend only on the BSA in DMEM but also on the concentration of the CTAB for nanorods suspended in stock solution. These gold nanorod aggregates were stable for more than one week. The aggregate size and hence its absorption resonance wavelength could be tuned up to the second biological therapeutic window easily. The presented method might be applied to get aggregates of any fixed shape and size of nanoparticles to obtain a large redshift in the absorption resonance wavelength tuning without changing the shape and size of the nanoparticles. It was also demonstrated that the aggregates could be optically heated using a broadband light source to get a desired increased temperature and there was no need for different laser light sources for nanoparticles and their aggregates separately. Use of broadband light has more advantages over laser light because it could cover a wide range of the absorption resonances of the nanorods and a single source could be employed for a variety of the nanoparticles and their combinations. After the photothermal interaction, the nanorods and aggregates were quite stable and showed almost the same resonance peak positions. The PCE of the gold nanorod aggregates were found almost equivalent or higher than the monodispersive form of the nanorods. The aggregation would enhance the photothermal heat generation also along with the higher redshift in the absorption spectra. The concentration of BSA and CTAB dependent large spectral redshift of absorption resonance and photothermal heat generation of the stable aggregates of small gold nanorods are the core results of the present study. The reported work might be used in therapeutics where the light penetration depth could be increased since the absorption resonance wavelength peaks could be varied towards higher wavelengths up to the second biological therapeutic window. The plasmonic coupling within the nanorods or nanoparticles in the aggregates would increase the sensitivity of imaging even for a single cell because aggregation would create a better hot-spot than the single nanoparticle. The higher PCE of the aggregates would be better for the antimicrobial application. When the nanoparticles are inserted in the body for \textit{in vivo} uses then there is a differential distribution of the nanoparticles. Somewhere distribution would be accumulated and sparse at other places. Our study shows that in such cases photothermal heat generation would be different also because of variable plasmonic coupling. Therefore, a broadband light source would be a better option for therapeutic uses.

\section*{Acknowledgement}
D.P. acknowledges the CSIR-India for the Nehru Science Postdoctoral Research Fellowship number HRDG/CSIR-Nehru PDF/EN, ES \& PS/EMR-I/04/2019. D.P. and S.S. acknowledge the support of the CSIR-CSIO Chandigarh for hosting the research.
\newpage
\vspace{2cm}
\begin{center}
{\Large \uppercase{Supporting Information}}
\end{center}
\section*{Estimation of CTAB concentration in the gold nanorod suspension}
In the process of gold nanorod synthesis, hexadecyltrimethylammonium bromide (CTAB) acts as a capping agent to stabilise the gold nanorod suspension~\cite{smith2008importance,shi2021strategies}. The concentration of CTAB was calculated using the elementary dilution formula $M_i \times V_i = M_f \times V_f$, where $M_{i,f}$ and $V_{i,f}$ were the molarity concentration of the CTAB and volume of the solution ($i = $ initial, $f = $ final). A similar approach has been utilised to calculate the CTAB concentration earlier~\cite{wu2015large}.

\begin{enumerate}

\item For preparing the seed solution, following salts and compounds were used: \\
9.75 ml CTAB (0.1 M),\\
0.25 ml HAuCl$_4$ (0.01M) and\\
0.6 ml NaBH$_4$ (0.01M).\\
The concentration of CTAB in the mixture of seed solution of a total 10.6 ml volume (V1) became 0.09198~M. 

\item For preparing the growth solution, the following salts and compounds were used: \\
90 ml CTAB (0.1 M),\\
5 ml HAuCl$_4$ (0.01M),\\
1 ml AgNO$_3$ (0.01M),\\
2 ml HCl (1 M) and\\
0.8 ml Ascorbic acid (0.1 M).\\
The concentration of CTAB in the mixture of growth solution of 98.8 ml volume (V2) became 0.09109 M. 

\item After adding the whole seed solution into the growth solution, the total volume (V3) became 109.4 ml and this volume had 0.09118~M CTAB concentration.

\item We used four round bottom tubes to centrifuge the nanorod suspensions. We added some amount of the DI water into V3 to make the solution up to 144 ml (V4). The molarity of CTAB in V4 was 0.06927 M.

\item The amount V4 was divided equally into all four round-bottom tubes. The first centrifuge was done at 15000 rounds per minute (rpm) and 30 $^{\circ}$C for 20 minutes. After the first centrifuge from each round bottom tube, a 34 ml supernatant was thrown and a 2 ml pellet was remaining in each tube. In this pellet, visually some nanorods diffused in the 2~ml volume and most of the nanorods settled down at the bottom of the tube. Since the nanorods synthesis happened in a homogeneous solution of CTAB, therefore, the concentration of CTAB in the 2~ml pellet was considered the same as for V4. Now 9 ml of DI water was added to each round bottom tube and the volume in each tube became 11 ml (V5). The volume V5 has 0.01259~M CTAB concentration. 
 
\item The V5 was centrifuged again at 15000 rpm and 30 $^{\circ}$C for 20 minutes. After the second centrifuge, 9 ml supernatant was removed from each tube and the remaining 2 ml pellet from each tube was put into a culture tube. In the nanorod kept culture tube, DI water was added to make the final volume 24 ml (V6). The V6 has 0.00420~M CTAB concentration. The V6 was the stock solution for further use. 

\item Diluting the V6 by ten times, we get the nanorod suspension with CTAB concentration C1 = 0.00042~M = 0.42~mM and this nanorod suspension was used for the first set of samples of nanorod aggregation. 

\item To get nanorod suspension of CTAB concentration C2 (< C1), a 2 ml of V6 was added in two 1.5 ml eppendorf tubes equally to balance and again centrifuged at 17000 rpm and 25 $^{\circ}$C for 20 minutes. After centrifuge, a visually transparent 0.8 ml supernatant was removed from each eppendorf tube and then 0.8 ml DI water was added to each tube. The final 2 ml volume (V7) was the stock solution of nanorods for further uses. The CTAB concentration in V7 was 0.00084 M. 

\item As in the previous case of C1, the V7 was also ten times diluted with DI water to get the nanorod suspension with CTAB concentration C2 = 0.00008~M = 0.08~mM and this suspension was used for the second set of nanorod aggregations experiments. 

\end{enumerate}
\bibliographystyle{ieeetr}
\bibliography{references}  
\end{document}